\documentclass[prd,onecolumn,notitlepage,amsmath,nofootinbib,superscriptaddress,showpacs,showkeys]{revtex4-1}

\usepackage[brazil,english]{babel}

\usepackage{amsmath}
\usepackage{graphicx}
\usepackage{amssymb}
\usepackage{esint}
\usepackage[usenames,dvipsnames]{color}

\usepackage{longtable}
\usepackage{dcolumn}
\usepackage{nicefrac}
\usepackage[normalem]{ulem}

\DeclareMathOperator{\cotanh}{cotanh}

\makeatother

\makeatletter

\begin{document}

\title{Quasinormal modes of generalized P\"{o}schl-Teller potentials}

\author{A. F. Cardona}
\email{andresfcj@usp.br}
\affiliation{Instituto de F\'{\i}sica, Universidade de S\~{a}o Paulo \\
C.P. 66318, 05315-970, S\~{a}o Paulo-SP, Brazil}

\author{C. Molina}
\email{cmolina@usp.br}
\affiliation{Escola de Artes, Ci\^{e}ncias e Humanidades, Universidade de
  S\~{a}o Paulo\\ Av. Arlindo Bettio 1000, CEP 03828-000, S\~{a}o
  Paulo-SP, Brazil}

\begin{abstract}

Using algebraic techniques we obtain quasinormal modes and frequencies associated to generalized forms of the scattering P\"{o}schl-Teller potential. This approach is based on the association of the corresponding equations of motion with Casimir invariants of differential representations of the Lie algebra $\mathfrak{sl}(2)$. 
In the presented development, highest weight representations are constructed and fundamental states are calculated. An infinite tower of quasinormal mode solutions is obtained by the action of a lowering operator.
The algebraic results are used in the analysis of the Cauchy initial value problem associated to the generalized P\"{o}schl-Teller potentials. For the scattering potentials considered, there are no late-time tails and the dynamics is always stable.

\end{abstract}


\maketitle

\section{Introduction}

The P\"{o}schl-Teller potential was originally introduced as a
potential for which the Schr\"{o}dinger equation is exactly solvable
\cite{Poschl:1933zz}.
In Quantum Mechanics, the work of P\"{o}schl and Teller was used to model physically relevant potential wells, as seen for example in \cite{PhysRev.42.210}.
More recently, an effective potential based on the P\"{o}schl-Teller expression
was considered in scattering problems. In this context, the term ``P\"{o}schl-Teller potential'' usually refers to the function
\begin{equation}
V(x) = \frac{V_{+}}{\cosh^{2} (\kappa x)} \,\, ,
\label{poschl-teller_def}
\end{equation}
with $V_{+} > 0$ and $\kappa > 0$, defined on the whole real line.
The potential~(\ref{poschl-teller_def}) is particularly important in gravitational physics, where it has been used to model the dynamics of fields propagating around compact objects such as black holes and wormholes (see for example \cite{Ferrari:1984zz,Cardoso:2003sw,Molina:2003ff,Molina:2003dc,Jing:2003wq,Konoplya:2005et,Molina:2012ay,Liu:2012zl,Fernando:2016ftj,Gonzalez:2017shu}).

There are no bound states associated to $V(x)$ in equation~\eqref{poschl-teller_def} \cite{Beyer:1998nu}, and the typical problem with this effective potential is not the one originally addressed by P\"{o}schl and Teller in \cite{Poschl:1933zz}. 
In scattering problems, one is frequently interested in quasinormal modes (QNMs), which are solutions of the relevant equations of motion with purely ingoing and outgoing boundary conditions. 
Related to a given quasinormal mode, there is a quasinormal frequency
that is usually complex-valued. Quasinormal spectra are relevant since they do not depend on the details of the initial conditions, but rather on the parameters of the background geometry.
They can be used in the construction of the Green function associated to the relevant equations of motion, and therefore they determine at least part of the dynamics in the considered systems.
A vast literature is concerned with the analysis of the issue 
(see for example the topical reviews \cite{Nollert:1999ji, kokkotas1999quasi, berti2009quasinormal, Konoplya:2011qq} and references therein).

The actual calculation of quasinormal modes and frequencies is usually a challenging mathematical problem, and only in few specific scenarios it can be done analytically. Among the techniques used we point out the algebraic approach, associating the relevant equation of motions with a convenient algebra
\cite{Frank:1984qd, Alhassid:1984uy, Wu:1989qoa, Cho, castro2010hidden, chen2010hidden,bertini2012conformal,chen2011quasi,kim2013quasinormal}.
Even though this idea is not new,%
\footnote{See for example \cite{Frank:1984qd, Alhassid:1984uy, Wu:1989qoa},
where the time independent version of the Schr\"{o}dinger equations with several potentials is treated.}
there has been recent attention in the algebraic treatment of various types of field perturbations on several background geometries. 
In this context, it is often said that the field equation possesses an underlying hidden symmetry, not related to the background isometries
\cite{castro2010hidden,chen2010hidden,bertini2012conformal,chen2011quasi,kim2013quasinormal}.

In the present work, we consider generalizations of the  P\"{o}schl-Teller potential~(\ref{poschl-teller_def}). One possible route would be toward the Eckart potential, which was first introduced for the study of electron tunneling \cite{eckart1930penetration}, but has also found application in gravitational physics in the study of the dynamics of Schwarzschild black hole~\cite{blome1984quasi} and Schwarzschild-de Sitter black hole~\cite{skakala2011quasi}.
We follow a different path, using scattering versions of the original (potential well) P\"{o}schl-Teller expressions.
More specifically, we study the relation between the scalar field equation with generalized forms of the P\"{o}schl-Teller potential and differential representations of the Lie algebra $\mathfrak{sl}(2)$. We find a direct relation between the equations of motion
and an invariant of the algebra, namely the Casimir element, allowing us to obtain solutions in a closed
form by means of a highest weight representation. The solutions are found to represent quasinormal modes, and the characteristics of the quasinormal spectra are explored.

The structure of the paper is as follows. In section~\ref{Equations} we present a generalized P\"{o}schl-Teller potential and introduce the pertinent equation of motion.
In section~\ref{algebra} we construct  representations of the 
$\mathfrak{sl}(2)$ algebra that will be used to 
solve the wave equations. 
In section~\ref{modes}, algebraic tools are used in the computation
of the quasinormal modes and frequencies.
A further generalization of the scattering P\"{o}schl-Teller potential is considered in section~\ref{further}, and the associated Cauchy initial value problem is discussed in section~\ref{cauchy}.
Final comments are presented in section~\ref{conclusion}, and concrete physical scenarios are discussed in the appendices.

\section{Generalized P\"{o}schl-Teller potentials}
\label{Equations}

In this work we study the following second order differential
equation defining the dynamics of a function $\Psi(t,x)$, 
\begin{equation}
\label{maineq}
-\frac{\partial^{2}\Psi(t,x)}{\partial t^{2}} + \frac{\partial^{2}\Psi(t,x)}{\partial x^{2}} 
= V(x)  \Psi(t,x) \,,
\end{equation}
with $t$ being a time variable and $x$ an spatial variable.
In the context of gravitational physics, the wave equation \eqref{maineq} appears in the study of scalar, vector and tensor perturbations on spherically symmetric and static geometries \cite{Nollert:1999ji, kokkotas1999quasi, Konoplya:2011qq}. 
The potential $V(x)$ depends on the characteristics of the background geometry.

We focus initially on generalized P\"{o}schl-Teller potentials with the form 
\begin{equation}\label{generalized_Poschl-Teller}
V(x) = \frac{V_{0}}{\cosh^{2}(\kappa x + \alpha)} \,.
\end{equation}
The parameter $\alpha$ is allowed to be complex.
Naturally, a particular case of $V(x)$ in equation~\eqref{generalized_Poschl-Teller} is the usual P\"{o}schl-Teller potential in equation~\eqref{poschl-teller_def}, obtained by setting $V_{0} = V_{+} > 0$, $\kappa > 0$ and $\alpha = 0$. In gravitational physics, this case appears most notably in the study of the dynamics of a scalar field on de Sitter black holes in the near extremal limit
\cite{Cardoso:2003sw,Molina:2003ff,Molina:2003dc,Jing:2003wq, Konoplya:2005et, Molina:2012ay}.

In addition to~(\ref{poschl-teller_def}), other interesting cases are possible if quasinormal modes are to be considered.%
\footnote{The original potential well discussed by P\"{o}schl and Teller \cite{Poschl:1933zz} can be obtained with functions constructed with $V_{0} > 0$, $\kappa = i k$, $\alpha = 0$; or with $V_{0} < 0$, $\kappa = i k$, $\alpha = i \pi/2$.} 
For example, with $V_{0} = - V_{-} < 0$, $\kappa > 0$ and $\alpha = i \pi/2$ we have
\begin{equation}
V(x) = \frac{V_{-}}{\sinh^{2} (\kappa x)} \,\, ,
\label{poschl-teller_sinh}
\end{equation}
where the domain of $V(x)$ in this case is defined to be the half real line,%
\footnote{
In the present work, the choice of $(-\infty,0)$ as the domain of the effective potentials in equations~\eqref{poschl-teller_sinh} and \eqref{full_Poschl_teller} is a matter of convention. This choice is consistent with the usual definition of the ''tortoise coordinate'' in AdS spacetimes \cite{Chan:1997,moss2002gravitational,Wang:2004bv}. But other characterizations for the open half real line, for example $(0,+\infty)$, could be equivalently used.
}
$x \in (-\infty,0)$. The modified form of P\"{o}schl-Teller effective potential in equation~\eqref{poschl-teller_sinh} is relevant when the gravitational perturbative dynamics on anti-de Sitter or de Sitter spacetimes is treated \cite{Du:2004jt, Ishibashi:2004wx}.

We are mainly interested in solutions of the scalar wave equation~\eqref{maineq} with the form
\begin{equation}
\Psi(t,x) = \psi(x) e^{-i \omega t} \,,
\label{exp-time}
\end{equation}
where $\omega$ is extended to the complex plane. With the sign convention in the argument of the exponential in equation~\eqref{exp-time}, stable solutions are those with $\textrm{Im}(\omega)<0$.
Considering wave functions with the time-dependence in equation~\eqref{exp-time}, we obtain the so-called time-independent version of the equation of motion:
\begin{equation}
\label{time_independent_eq}
\frac{d\psi(x)} {dx^{2}} 
+ \left[\omega^{2} -  V(x) \right] \psi(x) = 0 \,.
\end{equation}
It should be noticed that equation~\eqref{time_independent_eq} has the same form of the time-independent version of Schr\"{o}dinger equation, a relevant point when the original work of P\"{o}schl and Teller is considered \cite{Poschl:1933zz}.

Quasinormal modes and frequencies are especially important in scattering problems. 
The standard definition for quasinormal modes involves potentials $V(x)$ defined on the whole real line, $x \in (-\infty, +\infty)$. In that case, QNMs are solutions of equation~\eqref{time_independent_eq} behaving as purely ingoing waves in the limit $x \to -\infty$ and as purely outgoing waves as $x \to \infty$:
\begin{equation}
\label{boundaryQNM}
\psi(x) \sim \left\lbrace
\begin{array}{ll}
e^{-i\omega{x}} &\mbox{ as } x \to -\infty \\
e^{i\omega{x}} &\mbox{ as } x \to \infty
\end{array}
\right.  \, .
\end{equation}
The complex numbers $\omega \in \{\omega_{n} \}$ for which the boundary conditions~\eqref{boundaryQNM} are simultaneously satisfied are the so-called quasinormal frequencies \cite{Nollert:1999ji,kokkotas1999quasi,Konoplya:2011qq}.

When the effective potentials $V(x)$ are defined only on the half real line, that is $x \in (-\infty,0)$, it is usually introduced a modified version of the QNM boundary conditions, which can be expressed as
\begin{equation}
\label{boundaryQNM_modified}
 \psi(x) \sim \left\lbrace
 \begin{array}{ll}
 e^{-i\omega x} & \mbox{ as } x\to -\infty \\
 0 & \mbox{ as } x\to 0
 \end{array}
\right. \,.
\end{equation}
This version of the quasinormal mode definition will be relevant when the modified P\"{o}schl-Teller potential in equation~\eqref{poschl-teller_sinh} is considered, as we will see in the following sections.

\section{Algebraic description}
\label{algebra}

The equation of motion~\eqref{maineq} with the generalized P\"{o}schl-Teller potential is invariant
under time translations. As we shall see, another symmetry of this system is described by the Lie algebra $\mathfrak{sl}(2)$,
the algebra of $2 \times 2$ traceless matrices. A basis 
$\{H, E, F\}$ of $\mathfrak{sl}(2)$ satisfies the following relations:
\begin{equation}\label{algebracommutator}
[H,E] = 2E \,, \quad [H,F] = - 2F \,, \quad  [E,F] = H \,. 
\end{equation}

We aim to establish a connection between the 
equations of motion and a linear representation of the algebra 
$\mathfrak{sl}(2)$ in terms of differential
operator acting on the set of differentiable functions $\mathbb{R} \times \mathbb{R} \rightarrow \mathbb{C}$. 
Linearity reduces the problem to finding a mapping $\rho$ in terms of  
$\{\partial_{t}, \partial_{x}\}$, from the basis elements
$\{H, E, F\}$ to a set of operators 
$\{ \hat{L}_{0}, \hat{L}_{+}, \hat{L}_{-} \}$, 
\begin{equation}
 \rho(H) = \hat{L}_{0} \,, \quad \rho(E) = \hat{L}_{+} \,, \quad \rho(F) = \hat{L}_{-} \,,
\end{equation}
preserving the Lie bracket structure in equation~\eqref{algebracommutator},
\begin{equation}
\label{repr_bracket}
[\hat{L}_{0},\hat{L}_{+}] = 2  \hat{L}_{+} \, , \quad
[\hat{L}_{0},\hat{L}_{-}] = -2 \hat{L}_{-} \, , \quad
[\hat{L}_{+},\hat{L}_{-}] = \hat{L}_{0} \, . 
\end{equation}

We propose the following representation of $\mathfrak{sl}(2)$, which will be the core of the algebraic analysis developed in the present work:
\begin{align}
\label{cartan_cosh}
\hat{L}^{}_{0} &= \frac{2}{\kappa}\frac{\partial}{\partial{t}}\,,\\
\label{Lplus_cosh}
\hat{L}^{}_{+} &= \frac{1}{\kappa}e^{{\kappa}{t}}\left[-\sinh(\kappa x + \alpha)\frac{\partial}{\partial{t}}
- \cosh(\kappa x + \alpha)\frac{\partial}{\partial{x}}\right] \,, \\
\label{Lminus_cosh}
\hat{L}^{}_{-} &= \frac{1}{\kappa}e^{-{\kappa}{t}}\left[-\sinh(\kappa x + \alpha)\frac{\partial}{\partial{t}}
+ \cosh(\kappa x + \alpha)\frac{\partial}{\partial{x}}\right] \,.
\end{align}
It is straightforward to verify that $\hat{L}_{0}$, $\hat{L}_{+}$ and $\hat{L}_{-}$ in equations~\eqref{cartan_cosh}, \eqref{Lplus_cosh} and \eqref{Lminus_cosh} satisfy the algebraic relations~\eqref{repr_bracket}. It should be noted that the operators $\hat{L}^{}_{0}$, $\hat{L}^{}_{+}$ and $\hat{L}^{}_{-}$ cannot be diagonalized under the same basis.

Since quasinormal modes have the time dependence indicated in equation~\eqref{exp-time}, we select $\hat{L}_{0}$ as our diagonalizable operator. In representation theory language, $\hat{L}_{0}$ is chosen to be a Cartan operator.
In this way, a quasinormal mode must be an eigenvalue of $\hat{L}_{0}$,
\begin{equation}
 \hat{L}_{0} \left[ \psi(x) e^{-i \omega t} \right] = -i \frac{2\omega}{\kappa} \left[ \psi(x) e^{-i \omega t} \right] \,.
\label{L0_action_QNM}
\end{equation}

The Casimir invariant of the representation will be important in the development. The Casimir is an operator which commutes with all the operators associated to the basis in equations~\eqref{cartan_cosh}-\eqref{Lminus_cosh}, being given by
\begin{equation}
\label{sl2Rcasimir}
\hat{L}^{2} = \frac{1}{2} \, \hat{L}_{0} \, \hat{L}_{0} + \hat{L}_{-} \, \hat{L}_{+} + \hat{L}_{+} \, \hat{L}_{-} \,.
\end{equation}
In the present case, from equation~\eqref{sl2Rcasimir}, the Casimir operator associated with the representation in equations~\eqref{cartan_cosh}-\eqref{Lminus_cosh} is 
\begin{equation}
\label{casimirPoschTeller}
\hat{L}^{2} = - \frac{2}{\kappa^{2}}{\cosh^{2}(\kappa x + \alpha)}
\left(
- \frac{ \partial^{2}}{\partial t^{2}} 
+ \frac{ \partial^{2}}{\partial x^{2}} 
\right) \,.
\end{equation}

It follows from result~\eqref{casimirPoschTeller} that the equation of motion~\eqref{maineq} can be written as a constraint in the proposed representation of the algebra $\mathfrak{sl}(2)$:
\begin{equation}
\label{relation_Casimir_V-}
\hat{L}^{2} \Psi(t,x) = -2\frac{V_{0}}{\kappa^{2}}\Psi(t,x) \,.
\end{equation}
Thus, the value of the Casimir is directly related with the potential.

In the next section, we will use the algebraic description of the equation of motion for the determination of the quasinormal modes and frequencies derived from the generalized P\"{o}schl-Teller potential.

\section{Quasinormal modes and frequencies}
\label{modes}

\subsection{General results}

To obtain the solution of the equations~\eqref{L0_action_QNM} and \eqref{relation_Casimir_V-}, we will consider what is known as a highest weight representation \cite{fuchs2003symmetries,georgi1982lie}.
The solutions obtained with this method will be a set on functions $\{\Psi_{n}\}$ on which
the elements of the representation \eqref{repr_bracket} act as linear operators.
Since none of the elements of \eqref{repr_bracket} commute with each other,
only one operator can be chosen as acting diagonally on $\{\Psi_{n}\}$,
which by custom it is chosen to be $\hat{L}^{}_{0}$. 
A function $\Psi_{n}$ is said to be a weight function of weight $\lambda$ if 
\begin{equation}
\hat{L}_{0} \Psi_{n}  = \lambda \Psi_{n} \,.
\end{equation}

Having selected a diagonalizable operator on the representation, our solutions will be eigenvalues of both $\hat{L}_{0}$ and $\hat{L}^{2}$.
The functions $\{\Psi_{n}\}$ will not be eigenvalues of $\hat{L}_{\pm}$. Instead,
when acting on a function $\Psi$ they will increase or decrease the eigenvalue 
of the corresponding function with respect to $\hat{L}^{}_{0}$:
\begin{equation}
\hat{L}_{0}(\hat{L}_{\pm} \Psi)  = (\lambda \pm 2) (\hat{L}_{\pm}\Psi) \,.
\end{equation}
In that sense, it is said that $\hat{L}_{\pm}$ act as raising/lowering operators, and 
the whole set of functions can be obtained from the action of these operators on the 
weight space. 
For our purposes, we are considering an infinite dimensional representation but with a fundamental state.

In a highest weight representation, it is introduced a function $\Psi^{(0)}(t,r)$ satisfying the highest weight conditions,
\begin{align}
\hat{L}_{0} \Psi^{(0)}(t,x) & = h \Psi^{(0)}(t,x) \,, \label{highestweight_C1} \\
\hat{L}_{+} \Psi^{(0)}(t,x) & = 0 \,, \label{highestweight_C2}
\end{align}
where $h$ is the highest weight. From this fundamental state, an infinite 
number of solutions is obtained by successive applications of the lowering operator $\hat{L}_{-}$.

Since the action of $\hat{L}_{0}$ on a quasinormal mode is given by equation~\eqref{L0_action_QNM}, it follows that the quasinormal frequency $\omega_{0}$ associated to $\Psi^{(0)}$ is related to the constant $h$ as
\begin{equation}\label{omega_h}
\omega_{0} = i \frac{\kappa h}{2} \,.
\end{equation}
In terms of the highest weight $h$, the action of the Casimir $\hat{L}^{2}$ on $\Psi^{(0)}(t,r)$ is given by
\begin{equation}\label{casimirhighestweight}
\hat{L}^{2} \Psi^{(0)}(t,x) = 
\left( \frac{h^{2}}{2} + h \right) \Psi^{(0)}(t,x) \,.
\end{equation}
Direct comparison between equations~\eqref{relation_Casimir_V-} and \eqref{casimirhighestweight} allows one to solve $h$ in terms of $V_{0}$ and $\kappa$, 
\begin{equation}\label{weight_cosh}
h = -1 \pm 2i \, \sqrt{\frac{V_{0}}{\kappa^{2}} - \frac{1}{4}} \,.
\end{equation}
Using equations~\eqref{omega_h} and \eqref{weight_cosh}, the fundamental frequency $\omega_{0}$ can now be expressed in terms of the potential parameters $\kappa$ and $V_{0}$ as
\begin{equation}\label{first_freq}
\omega_{0} = \kappa\left( -i\frac{1}{2} \pm \sqrt{\frac{V_{0}}{\kappa^{2}} - \frac{1}{4}} \right)  \,.
\end{equation}

To find the explicit form of the corresponding fundamental mode $\Psi^{(0)}(t,r)$ associated to $\omega_{0}$, we use the highest weight conditions in equations~\eqref{highestweight_C1} and \eqref{highestweight_C2}, which are translated to the following differential equations:
\begin{equation}
 \frac{2}{\kappa} \frac{\partial \Psi^{(0)}(t,x)}{\partial t} = h \Psi^{(0)}(t,x) \, ,
\label{L0_eq_cosh}
\end{equation}
\begin{equation}
-\sinh( \kappa x + \alpha )\frac{\partial \Psi^{(0)}(t,x)}{\partial{t}} 
- \cosh(\kappa x + \alpha)\frac{\partial \Psi^{(0)}(t,x)}{\partial x} = 0 \, .
\label{Lp_eq_cosh}
\end{equation}
Equations~\eqref{L0_eq_cosh} and \eqref{Lp_eq_cosh} can be solved exactly, with
\begin{equation}\label{HW_function_geral}
\Psi^{(0)}(t, x) =
A e^{- i \omega_{0} t} \cosh(\kappa x + \alpha )^{i \omega_{0}/\kappa} \,,
\end{equation}
where $A$ is an integration constant. 

The complete spectrum $\{ \omega_{n} \}$ can be expressed in a closed form. For this purpose, we use the following property,
\begin{equation}\label{induction}
 [\hat{L}_{0},(\hat{L}_{\pm})^{n}] = \pm 2n (\hat{L}_{\pm})^{n} \, ,
\end{equation}
which can be proved by induction. In equation~\eqref{induction}, $n$ is a non-negative integer.
The frequency associated to the mode $\Psi^{(n)}(t,x)$ is given by the action of the operator $\hat{L}_{0}$. Using equation~\eqref{induction}, we obtain that
\begin{equation}
\hat{L}_{0} \Psi^{(n)}(t,x) = 
\left( 
[ \hat{L}_{0} , \hat{L}_{-}^{n}] + \hat{L}_{-}^{n} \hat{L}_{0} 
\right) \Psi^{(0)}(t,x) 
= \left(h - 2n \right) \Psi^{(n)}(t,x) \,.
\label{L0Psi}
\end{equation}
With equation~\eqref{L0Psi}, the fundamental and overtone frequencies are given by
\begin{equation}
\omega_{n} =  i \frac{\kappa}{2}(h - 2n) 
= \kappa \left[ - i\left(n + \frac{1}{2}\right) \pm \sqrt{\frac{V_{0}}{\kappa^{2}} - \frac{1}{4}} \right] \, , \,\,
n = 0,1,2\ldots \, .
\label{qnf}
\end{equation}

Higher order solutions can be obtained by the successive application of the operator $\hat{L}_{-}$ to the fundamental mode $\Psi^{(0)}(t,x)$, 
\begin{equation}\label{n-mode-geral}
\Psi^{(n)}(t,x) = (\hat{L}_{-})^{n} \Psi^{(0)}(t,x) \,,
\end{equation}
and there will be an infinite number of them. For instance, considering the second mode, one has
\begin{equation}
\Psi^{(1)}(t,x) = 
A e^{- i \omega_{1} t} 
\sinh(\kappa x + \alpha) \cosh(\kappa x + \alpha)^{i\frac{\omega_{0}}{\kappa}} \, .
\end{equation}
For the third mode, it is obtained that
\begin{equation}
\Psi^{(2)}(t,x) = 
A e^{- i \omega_{2} t} \, 
\cosh(\kappa x + \alpha)^{i\frac{\omega_{0}}{\kappa}} 
 \left[
\cosh(\kappa x + \alpha) + \left( 1 + h \right) \sinh^{2} (\kappa x + \alpha)
\right] \, ,
\end{equation}
and so on.

\subsection{Boundary conditions}

A boundary condition analysis is essential in the characterization
of quasinormal modes. Besides being solutions of the wave equation~\eqref{maineq},
they must satisfy the appropriate boundary conditions~\eqref{boundaryQNM} or \eqref{boundaryQNM_modified}.
In the present work, we are interested in real effective potentials.
With this requirement, we can restrict our analysis to only two particular
cases of the complex parameter $\alpha$, 
namely $\alpha=0$ and $\alpha=i\,\pi/2$.
In addition, since we focus on the quasinormal problem, the
potential must not be negative-definite, and therefore we obtain the
particular cases presented in equations~\eqref{poschl-teller_def} and \eqref{poschl-teller_sinh}. For other values
of $\alpha$, we have effective potentials that can be reduced to~\eqref{poschl-teller_def} or \eqref{poschl-teller_sinh} by real translations in the variable $x$.

Let us consider the case $\alpha=0$ and $V_{0}=V_{+}>0$ in the development from the previous section. This case corresponds to the usual P\"{o}schl-Teller potential \eqref{poschl-teller_def}, proportional to $\cosh^{-2} (\kappa x)$.
This is the well-known setup, but we will review it considering the algebraic approach introduced here.
For latter convenience, we will introduce some new notation. We denote the wave functions with $\alpha = 0$ as $\psi^{(n)}(t,x)$. The basis operators will be denoted $\{ \hat{P}_{0}, \hat{P}_{+}, \hat{P}_{-} \}$ for $\alpha = 0$.
Explicitly, we have for the fundamental mode
\begin{equation}
\psi^{(0)}(t, x) =
A e^{- i \omega_{0} t} \cosh(\kappa x)^{i \omega_{0}/\kappa} \,,
\label{HW_function_cosh}
\end{equation}
and for the higher overtones,
\begin{equation}
\psi^{(n)}(t,x) = (\hat{P}_{-})^{n} \psi^{(0)}(t,x) \,.
\label{n-mode-cosh}
\end{equation}
The frequencies have always non-null imaginary components:
\begin{equation}
\omega_{n}  =  
\kappa \left[ - i\left(n + \frac{1}{2}\right) \pm \sqrt{\frac{V_{+}}{\kappa^{2}} - \frac{1}{4}} \right] \, , \,\, n = 0,1,2, \ldots \,\, .
\label{spectrum1}
\end{equation}

Considering the limits $x \to \pm\infty$ for the solutions in equation~\eqref{HW_function_cosh}, one verifies that the functions $\psi^{(n)}(t,x)$  have the correct quasinormal mode asymptotic behavior
\begin{equation}
 \psi^{(n)}(t,x) \sim \left\lbrace
 \begin{array}{ll}
 e^{i\omega x} & \mbox{ as } x \to \infty \\
  e^{-i\omega x} & \mbox{ as } x\to -\infty
 \end{array}
\right. \,,
\end{equation}
matching the standard quasinormal mode boundary conditions~\eqref{boundaryQNM}.

For the case $\alpha=i\,\pi/2$ and $V_{0}=-V_{-}<0$, we obtain
the modified P\"{o}schl-Teller potential in equation~\eqref{poschl-teller_sinh}, proportional to $\sinh^{-2} (\kappa x)$.
Also for latter convenience, we will denote the wave functions with $\alpha =  i \pi/2$ as $\varphi^{(n)}(t,x)$. The basis operators will be denoted $\{ \hat{M}_{0}, \hat{M}_{+}, \hat{M}_{-} \}$ when $\alpha = i \pi/2$. For this particular representation the Casimir operator is given by
\begin{equation}
\label{casimirPoschTeller_sinh}
\hat{M}^{2} = \frac{2}{\kappa^{2}}{\sinh^{2}(\kappa x)}
\left(
- \frac{ \partial^{2}}{\partial t^{2}} 
+ \frac{ \partial^{2}}{\partial x^{2}} 
\right) \,.
\end{equation}

General results from previous subsection give us
\begin{equation}
\varphi^{(0)}(t, x) =
A e^{-i\omega_0 t}\sinh(\kappa x)^{i\frac{\omega_{0}}{\kappa}} 
\label{HW_function_sinh}
\end{equation}
for the fundamental mode and
\begin{equation}
\varphi^{(n)}(t,x) = (\hat{M}_{-})^{n} \varphi^{(0)}(t,x)
\label{n-mode-sinh}
\end{equation}
for the higher overtones. As an important characteristic of this case, the spectrum turns out to be purely imaginary. We have two non-equivalent sets of solutions of the wave equation, $(+)$ and $(-)$, characterized by the frequencies
\begin{equation}
i\kappa\left[ -\left(n + \frac{1}{2}\right) \pm \sqrt{\frac{V_{-}}{\kappa^{2}} +\frac{1}{4} } \right] 
\,\, , \, n = 0,1,2, \ldots \,\, .
\end{equation}
The $(-)$ solutions are necessarily stable. But the $(+)$ solutions could describe unstable modes, with positive imaginary parts (for large enough values of $V_{-}/\kappa^{2}$). We will show in the following that only stable solutions are quasinormal modes.

Let us consider the boundary conditions to be satisfied. Functions $\{ \varphi^{(n)}(t,x) \}$ in equations~\eqref{HW_function_sinh}-\eqref{n-mode-sinh} are defined only on the half real line, $x \in (-\infty,0)$, and therefore they cannot satisfy the standard quasinormal boundary condition~\eqref{boundaryQNM}. But they could satisfy the modified quasinormal mode conditions in  \eqref{boundaryQNM_modified}. 
On the other hand, only the $(-)$ solutions satisfy the Dirichlet boundary condition prescribed in \eqref{boundaryQNM_modified}. That is, only for those solutions we have 
\begin{equation}
 \lim_{x \to 0^{-}} \varphi^{(n)}(t,x) \rightarrow 0 \,,
\end{equation}
as can be straightforwardly verified. Therefore, the unstable $(+)$ solutions are not quasinormal modes. The quasinormal frequencies associated to the modified P\"{o}schl-Teller potential are then given by
\begin{equation}
\omega_{n} = 
- i\kappa\left[ \left(n + \frac{1}{2}\right) + \sqrt{\frac{V_{-}}{\kappa^{2}} +\frac{1}{4} } \right] 
\,\, , \, n = 0,1,2, \ldots \,\, .
\label{spectrum2}
\end{equation}

\section{Further generalization of the P\"{o}schl-Teller potential}
\label{further}

We consider now a further generalization of the P\"{o}schl-Teller potential presented in equation~\eqref{poschl-teller_def}, combining both terms $\cosh^{-2}(\kappa x)$ and $\sinh^{-2}(\kappa x)$ as
\begin{equation}
V(x) = \frac{V_{+}}{\cosh^{2} (\kappa x)} + \frac{V_{-}}{\sinh^{2} (\kappa x)} \,\, ,
\label{full_Poschl_teller}
\end{equation}
with $V_{+}>0$ and $V_{-}>0$. We have again a scattering problem, and motivations for considering this potential can be found in the study of the perturbative dynamics of different types of fields \cite{Du:2004jt,Ishibashi:2004wx,lopez2006quasinormal}.
Regarding the domain of the the full P\"{o}schl-Teller potential, 
$V(x)$ diverges in the limit $x \to 0$, since the term proportional
to $\sinh^{-2}(\kappa x)$ becomes dominant. It follows that the potential is defined on the half real line, $x \in (-\infty,0)$.

In sections~\ref{algebra} and \ref{modes}, we considered
two particular representations of the algebra $\mathfrak{sl}(2)$: \texttt{(Rep1)}, as the operators $\{ \hat{P}_{0}, \hat{P}_{+}, \hat{P}_{-} \}$, obtained setting $\alpha=0$ in equations~\eqref{cartan_cosh}-\eqref{Lminus_cosh}; and \texttt{(Rep2)}, as the operators $\{ \hat{M}_{0}, \hat{M}_{+}, \hat{M}_{-} \}$, obtained setting $\alpha=i \pi/2$ in equations~\eqref{cartan_cosh}-\eqref{Lminus_cosh}. 
Both representations \texttt{(Rep1)} and \texttt{(Rep2)} share the same Cartan operator $\hat{L}_{0} = \hat{P}_{0} = \hat{M}_{0}$. However, their Casimir operators, denoted by $\hat{P}^{2}$ and $\hat{M}^{2}$, 
\begin{eqnarray}
\hat{P}^{2} & = & \frac{1}{2} \, \hat{P}_{0} \, \hat{P}_{0} + \hat{P}_{-} \, \hat{P}_{+} + \hat{P}_{+} \, \hat{P}_{-} \,, \\
\hat{M}^{2} & = & \frac{1}{2} \, \hat{M}_{0} \, \hat{M}_{0} + \hat{M}_{-} \, \hat{M}_{+} + \hat{M}_{+} \, \hat{M}_{-} \,,
\end{eqnarray}
and given explicitly by expressions \eqref{casimirPoschTeller} and \eqref{casimirPoschTeller_sinh}, do not commute.
Hence the eigenvalues of one of the Casimir do not constitute solutions for the other case.
Nevertheless, a solution for the full P\"{o}schl-Teller potential in equation~\eqref{full_Poschl_teller} can be constructed from the particular solutions $\psi^{(n)}(t,x)$ and $\varphi^{(n)}(t,x)$, associated to the potentials in equations~\eqref{poschl-teller_def} and \eqref{poschl-teller_sinh} respectively.

We will denote the product of the highest weight solution of both representations in equations~\eqref{HW_function_cosh} and \eqref{HW_function_sinh} by $\Psi^{(0)}$.
One obtains that 
\begin{eqnarray}
\Psi^{(0)}(t,x) & = & \psi^{(0)}(t,x)\varphi^{(0)}(t,x) \nonumber \\
 & = & A\cosh(\kappa x)^{-\frac{1}{2}h_{+}}
   \sinh(\kappa x)^{-\frac{1}{2}h_{-}}
   \exp^{\frac{\kappa}{2}\left(h_{+} + h_{-}\right)t} \, ,
   \label{general_fund_mode}
\end{eqnarray}
where $h_{+}$ and $h_{-}$ are the highest weight constants of \texttt{(Rep1)} and \texttt{(Rep2)} respectively:
\begin{eqnarray}
\hat{P}_{0} \psi^{(0)}(t,x) & = & h_{+} \psi^{(0)}(t,x) \,, \\
\hat{M}_{0} \varphi^{(0)}(t,x) & = & h_{-} \varphi^{(0)}(t,x) \,.
\end{eqnarray}

In the following, it will be shown that $\Psi^{(0)}(t,x)$ is a solution of the wave equation with the full P\"{o}schl-Teller~\eqref{full_Poschl_teller} is considered. 
Acting with the D'Alambertian operator $(-\partial_{t}^{2} + \partial_{x}^{2})$ on $\Psi^{(0)}(t,x)$ and taking into account the relation between the Casimir elements and the height of the potentials, one has
\begin{equation}
\left(-\frac{\partial^{2}}{\partial t^{2}}
+ \frac{\partial^{2}}{\partial x^{2}} \right)
\Psi^{(0)}(t,x)
= V(x)\varphi^{(0)}\psi^{(0)}
+ 2\left[\varphi'^{(0)}\psi'^{(0)} - \dot{\varphi}^{(0)}\dot{\psi}^{(0)} \right]\,.
\label{equation_product}
\end{equation}
In equation~\eqref{equation_product}, $\varphi'$ denotes derivative with respect to $x$, and $\dot{\varphi}$ denotes derivative with respect to $t$.
Also, we can show that the additional term on the right hand side of equation~\eqref{equation_product} is identically zero given the properties of the highest weight functions. Indeed, from the highest weight conditions, we obtain the following equalities,
\begin{eqnarray}
\label{exp1}
\tanh(\kappa x)\frac{\partial}{\partial{t}}\psi^{(0)}(t,x) = 
\frac{\partial}{\partial x}\psi^{(0)}(t,x) \,, \\
\label{exp2}
\cotanh(\kappa x)\frac{\partial}{\partial{t}}\varphi^{(0)}(t,x) = 
\frac{\partial}{\partial x}\varphi^{(0)}(t,x) \,.
\end{eqnarray}
Taking the product of the expressions in equations~\eqref{exp1} and \eqref{exp2}, we have
\begin{equation}\label{identity}
\varphi'^{(0)} \psi'^{(0)} = \dot{\varphi}^{(0)}\dot{\psi}
^{(0)} \,,
\end{equation}
from which the second term in the right-hand side of equation~\eqref{equation_product} is canceled. Hence,
\begin{equation}
\left[-\frac{\partial^{2}}{\partial t^{2}}
+ \frac{\partial^{2}}{\partial x^{2}} \right]
\Psi^{(0)}(t,x)
= 
V(x) \Psi^{(0)}(t,x) \,,
\end{equation}
and therefore $\Psi^{(0)}=\varphi^{(0)} \psi^{(0)}$ solves the wave equation~\eqref{maineq} with the potential~\eqref{full_Poschl_teller}, as we wanted to show.

We can now obtain the fundamental quasinormal frequency for the mode $\Psi^{(0)}(t,x)$. The action of the Cartan operator $\hat{L}_{0} = \hat{P}_{0} = \hat{M}_{0}$ on $\Psi^{(0)}(t,x)$ is
\begin{equation}
\hat{L}_{0} \Psi^{(0)}(t,x) = (h_{+} + h_{-}) \Psi^{(0)}(t,x) \,.
\end{equation}
It follows that the fundamental frequency of $\Psi^{(0)}(t,x)$ is given by the sum of the frequencies of the fundamental modes associated to  $\hat{P}_{0}$ and $\hat{M}_{0}$:
\begin{equation}
\frac{\partial}{\partial{t}}\Psi^{(0)}(t,x) = \frac{\kappa}{2}(h_{+} + h_{-})
\Psi^{(0)}(t,x)
= -i\omega_0\Psi^{(0)}(t,x) \,,
\label{L_on_Psi0}
\end{equation}
with
\begin{equation}
\omega_{0} =  
\kappa\left[ 
\pm \sqrt{\frac{V_{+}}{\kappa^{2}} -\frac{1}{4}} 
- i\left(1 + \sqrt{\frac{V_{-}}{\kappa^{2}} + \frac{1}{4}} \right) 
\right] \,.
\label{omega0_full_1}
\end{equation}

It can be readily verified that the function $\Psi^{(0)}(t,x)$ in equation~\eqref{general_fund_mode} satisfies the modified boundary conditions presented in equation~\eqref{boundaryQNM_modified}.
Therefore, $\Psi^{(0)}(t,x)$ and the associated $\omega_{0}$ are proper quasinormal modes and frequencies of the full P\"{o}schl-Teller potential.

\section{Quasinormal modes and an initial value problem}
\label{cauchy}

Quasinormal solutions of the generalized P\"{o}schl-Teller potentials can also be investigated through the analysis of an associated initial value problem. In fact, quasinormal modes dominate the intermediate and (possibly) the late-time field evolution. The initial value problem approach will complement and corroborate the algebraic results already presented.

Let us consider the Cauchy initial value problem associated to the hyperbolic equation~\eqref{maineq}. In this formulation, initial data are given by two functions $F$ and $G$, where
\begin{equation}
\Psi(0,x) = F(x) \,\, , \,\, 
\frac{\partial\Psi}{\partial t} (0,x) = G(x) \, .
\end{equation}
Since we are interested in a quasinormal mode evolution, we will consider initial conditions with a sharp peak and fast decay \cite{Nollert:1999ji,kokkotas1999quasi}.
For most of the development presented here, the initial data have the form
\begin{equation}
F(x) = A_{1} \, e^{-\sigma_{1} x^{2}} \,\, , \,\,
G(x) = A_{2} \, e^{-\sigma_{2} x^{2}} \, .
\end{equation}

An issue to be considered is the eventual existence of a late-time tail. That tail, if it exists, would dominate the field decay for the late-time regime, supplanting the quasinormal mode phase. The problem can be analytically addressed considering the asymptotic form of the effective potential. Asymptotically, with $-x$ large, one obtains that
\begin{equation}
V(x) = 4 \left( V_{+} + V_{-}  \right) \, e^{-2\kappa x} 
+ o\left( e^{2\kappa x}  \right) \,.
\end{equation}
That is, the potential decreases like an exponential. It is then shown in \cite{Price,Ching} that a potential with this form does not generate tails. 

\begin{figure}[h]
\includegraphics[clip,width=0.9\linewidth]{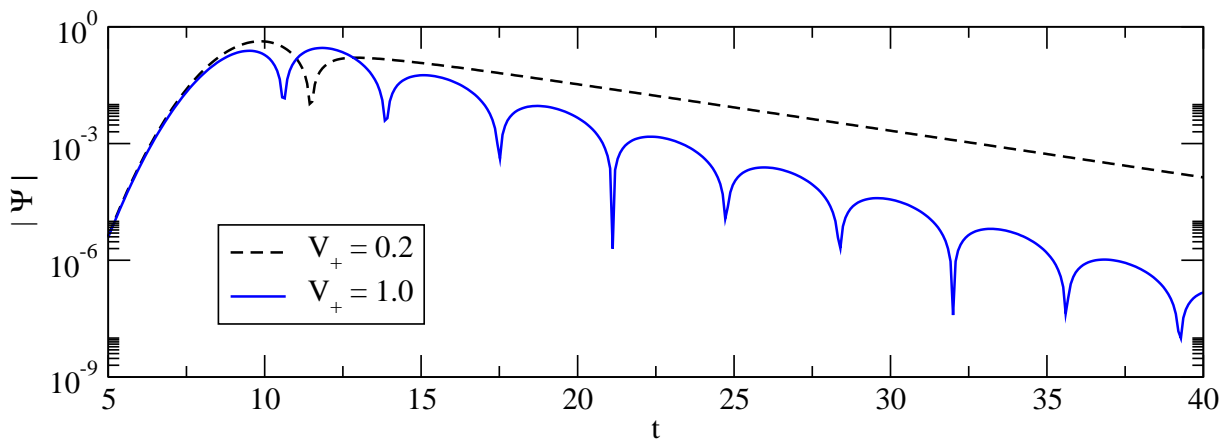}
\caption{Semi-log graphs for the field evolution with the usual P\"{o}schl-Teller potential. The oscillatory ($V_{+}/\kappa^{2} > 1/4$) and non-oscillatory ($0 < V_{+}/\kappa^{2} \le 1/4 $) regimes are shown. We used $\kappa=1$ and $V_{-}=0$.}

\label{usual-PT}
\end{figure}

\begin{figure}[h]
\includegraphics[clip,width=0.9\linewidth]{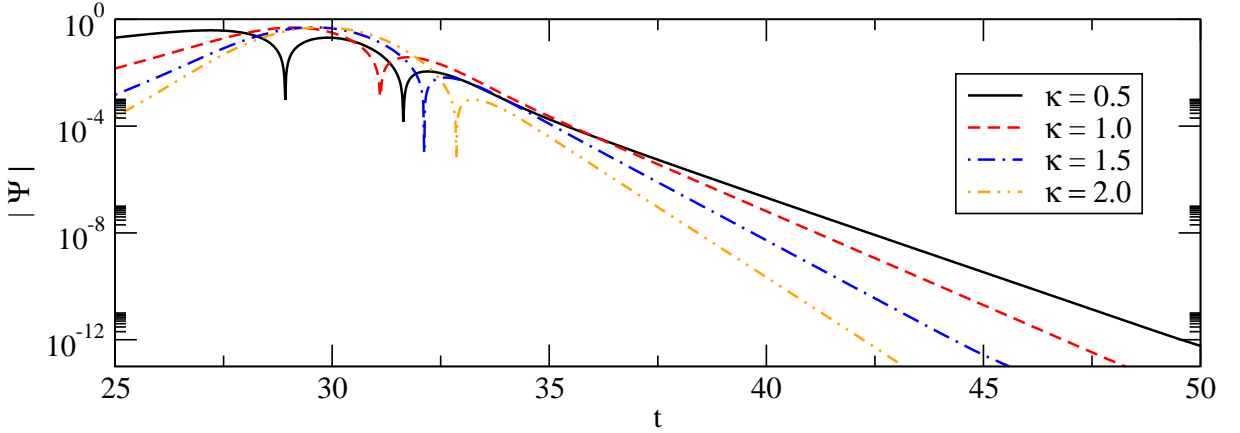}
\caption{Semi-log graphs for the field evolution with the modified P\"{o}schl-Teller potential. The late-time decay is always exponential and non-oscillatory. 
We used $V_{-}=1$ and $V_{+}=0$.}
\label{modified-PT}
\end{figure}

Therefore, we arrive at a qualitative description of the time evolution of the field at a fixed position. After an initial transient, which depends on the initial conditions, follows the quasinormal mode phase. The late-time field evolution is then dominated by the fundamental quasinormal mode. The field decay can be oscillatory or non-oscillatory, depending on the existence of a non-null real part in the fundamental quasinormal frequency.
The dynamics is always stable, that is, the function $\Psi(t,x)$ is bounded. These qualitative features of the field dynamics are illustrated in figures~\ref{usual-PT} and \ref{modified-PT}.

For a quantitative comparison between the analytic expressions found for the quasinormal frequencies and the results from the time evolution, we employ numerical techniques. We use an explicit finite difference scheme to numerically integrate the field equation~\eqref{maineq}. A discretized version of equation~\eqref{maineq} is obtained with
\begin{equation}
t \rightarrow t_{i}=t_{0}+i\,\Delta t \,\,\, , \,\,\, i=0,1,2,\ldots  \,,
\label{descreteT}
\end{equation}
\begin{equation}
x \rightarrow x_{j}=x_{0}+j\,\Delta x \,\,\, , \,\,\, j=0,1,2,\ldots   \,\,.
\label{descreteX}
\end{equation}
With this discretization, the wave equation is approximated by the difference equation
\begin{equation}
\psi_{N} = \left( 2 - \Delta t^{2}\, V_{C}\right) \,
\psi_{C} - \psi_{S} + \frac{\Delta t^{2}}{\Delta x^{2}}
\left(\psi_{E} - 2\psi_{C}+\psi_{W}\right) \,\, ,
\label{psiN}
\end{equation}
where
\begin{gather}
\Psi_{N} = \Psi\left(t_{i+1},x_{j}\right) \,\,\, , \,\,\,
\Psi_{E} = \Psi\left(t_{i},x_{j+1}\right) \,\, , \nonumber \\
\Psi_{C} = \Psi\left(t_{i},x_{j}\right) \,\,\, , \,\,\,
\Psi_{W} = \Psi\left(t_{i},x_{j-1}\right) \,\, , \nonumber \\
\Psi_{S} = \Psi\left(t_{i-1},x_{j}\right) \,\,\, , \,\,\,
V_{C} = V_{\ell} \left(t_{i},x_{j}\right) \,\, .
\end{gather}
With the described method for the numerical treatment of the Cauchy problem, we performed an extensive exploration on the parameter space of the generalized P\"{o}schl-Teller potentials. 

We first consider the usual P\"{o}schl-Teller potential ($V_{+}>0$ and $V_{-}=0$) defined on the whole real line. Although already carefully studied, one characteristic of the time evolution generated by the usual P\"{o}schl-Teller potential that was not stressed in the pertinent literature is the existence of oscillatory and non-oscillatory late-time field decay.%
\footnote{In the analysis of near extreme geometries, a non-oscillatory decay never appears because in those scenarios $V_{+}/\kappa^{2}>1/4$ \cite{Cardoso:2003sw,Molina:2003ff,Molina:2003dc,Jing:2003wq, Konoplya:2005et, Molina:2012ay}.} 
From the results of section~\ref{modes}, if $ V_{+}/\kappa^{2} > 1/4 $ and $V_{-}=0$, the real and imaginary parts of the fundamental frequencies are non-null, being given by
\begin{eqnarray}
\textrm{Re} \left(\omega_{0}^{\pm} \right) & = & \pm \kappa \, \sqrt{\frac{V_{+}}{\kappa^{2}}-\frac{1}{4}} \,, \\
\textrm{Im} \left(\omega_{0}^{\pm} \right) & = & -\frac{\kappa}{2} \,.
\end{eqnarray}
In this case, the sign choice is not relevant. The late-time decay is oscillatory and exponentially attenuated. This is the usual picture of the time evolution with the usual P\"{o}schl-Teller potential.
But in the regime where $0 < V_{+}/\kappa^{2} \le 1/4 $ and $V_{-}=0$, the fundamental quasinormal frequencies are purely imaginary:
\begin{eqnarray}
\textrm{Re} \left(\omega_{0}^{\pm} \right) & = & 0 \,, \\
\textrm{Im} \left( \omega_{0}^{+} \right) & = &
-\kappa\left(\frac{1}{2} +\sqrt{\frac{1}{4}-\frac{V_{+}}{\kappa^{2}}}\right) \,, \\
\textrm{Im} \left(\omega_{0}^{-} \right) & = &
- \kappa \left(\frac{1}{2} -\sqrt{\frac{1}{4}-\frac{V_{+}}{\kappa^{2}}}\right) \,.
\end{eqnarray}
Both $(\pm)$ modes are stable, but the $(-)$ mode has the lowest absolute value of its imaginary part, and therefore dominates the late-time decay. 
The existence of oscillatory and non-oscillatory modes for the usual P\"{o}schl-Teller potential can be readily seen in the numerical evolution. We illustrate this point in figure~\ref{usual-PT}.

Considering the modified P\"{o}schl-Teller potential ($V_{+}=0$ and $V_{-}>0$), it is apparent from the results of section~\ref{modes} that the late-time decay is always non-oscillatory, with an exponential coefficient given by
\begin{equation}
\textrm{Im} \left( \omega_{0} \right) = 
-\kappa\left(
\frac{1}{2} + \sqrt{\frac{V_{-}}{\kappa^{2}}+\frac{1}{4}}
\right) \,.
\end{equation}
We present typical results for the field evolution with the modified P\"{o}schl-Teller potential in figure~\ref{modified-PT}. From the data, the numerical evaluation for the fundamental frequency $\omega_{0}^{num}$ can be made. The comparison between the analytical and numerical results are presented in table~\ref{qnm-1}. The concordance of the two approaches is very good.

\begin{table}[h]
\caption{Analytical and numerical results for the fundamental quasinormal frequencies associated to the modified P\"{o}schl-Teller potential. The relative differences $\Delta\%$ between the results are also indicated. We use $V_{-}=1$ and $V_{+}=0$.}
\label{qnm-1}
\begin{tabular*}{\columnwidth}{*{4}{c@{\extracolsep{\fill}}}}
\hline
$\kappa$ & $\textrm{Im}(\omega_{0})$ & $\textrm{Im}(\omega_{0}^{num})$ &  $\Delta\%$ \\
\hline
0.25 & -1.1328 & -1.1607 & 2.46\\
0.50 & -1.2808   & -1.2965 & 1.23\\
1.00 & -1.6180   & -1.6196 & 0.10\\
1.50 & -2.0000   & -2.0018 & 0.09\\
2.00 & -2.4142   & -2.4174 & 0.13\\
\hline
\end{tabular*}
\end{table}

The field evolution with the full P\"{o}schl-Teller potential ($V_{+}>0$ and $V_{-}>0$) combines elements from the dynamics associated to the usual and modified P\"{o}schl-Teller potentials. 
We observe oscillatory and non-oscillatory regimes. Considering the results in section~\ref{further}, we observe that if $V_{+}/\kappa^{2} > 1/4$ the late-time decay is characterized by an oscillatory behavior,
\begin{eqnarray}
\textrm{Re} \left( \omega_{0}^{\pm} \right) & = & \pm \kappa \, \sqrt{\frac{V_{+}}{\kappa^{2}}-\frac{1}{4}}\,\,, \\
\textrm{Im} \left( \omega_{0}^{\pm} \right) & = & -\kappa\left(1+\sqrt{\frac{V_{-}}{\kappa^{2}}+\frac{1}{4}}\right) \,.
\end{eqnarray}
If $0<V_{+}/\kappa^{2} \le 1/4$, the field decays monotonically, with
\begin{eqnarray}
\textrm{Re} \left( \omega_{0}^{-} \right) & = & 0 \,, \\
\textrm{Im} \left( \omega_{0}^{-} \right) & = & -\kappa\left(1+\sqrt{\frac{V_{-}}{\kappa^{2}}+\frac{1}{4}}-\sqrt{\frac{1}{4}-\frac{V_{+}}{\kappa^{2}}}\right) \,.
\end{eqnarray}
The numerical results for the full P\"{o}schl-Teller potential are compared with the analytical formulas in table~\ref{qnm-2}, showing very good agreement between the two methods.

\begin{table}[h]
\caption{Analytical and numerical results for the fundamental quasinormal frequencies associated to the full P\"{o}schl-Teller potential.  The relative differences $\Delta\%$ between the results are also indicated. We use $V_{-}=1$ and $\kappa=1$.}
\label{qnm-2}
\begin{tabular*}{\columnwidth}{*{5}{c@{\extracolsep{\fill}}}}
\hline
$V_{+}$ & $\textrm{Re} \left( \omega_{0} \right)$ & 
$\textrm{Im} \left( \omega_{0} \right)$ &
$\textrm{Re} \left( \omega_{0}^{num} \right) (\Delta\% )$ &
$\textrm{Im} \left( \omega_{0}^{num} \right) (\Delta\% )$ \\
\hline
0 & 0 & -1.6180 & 0  & -1.6204 (0.15)\\
0.05 & 0 & -1.6708 & 0 & -1.6750 (0.25)\\
0.1 & 0 & -1.7307 & 0 & -1.7386 (0.46)\\
0.2 & 0 & -1.8944 & 0 & -1.9352 (2.15)\\
0.5 & 0.5000 & -2.1180 & 0.5061 (1.21) & -2.1288 (0.51)\\
1.0 & 0.8660 & -2.1180 & 0.8664 (0.05) & -2.1190 (0.05) \\
2.0 & 1.3229 & -2.1180 & 1.3243 (0.11) & -2.1003 (0.84)\\
5.0 & 2.1794 & -2.1180 & 2.1634 (0.74) & -2.1139 (0.19)\\
10.0 & 3.1225 & -2.1180 & 3.1145 (0.26) & -2.1305 (0.59)\\
\hline 
\end{tabular*}
\end{table}

\section{Final comments}
\label{conclusion}

In the present work, we introduced generalizations of the scattering P\"{o}schl-Teller potential in equation~\eqref{poschl-teller_def}. Quasinormal modes and frequencies were obtained for the considered effective potentials and their role in the associated Cauchy initial value problem was discussed.
The approach presented here was mainly algebraic, which stressed the underling hidden symmetries of the wave equation. 
The P\"{o}schl-Teller-type potentials considered were shown to be associated to a representation of the algebra $\mathfrak{sl}(2)$. 
The method is based on highest weight representations, in which there is a fundamental state from which an infinite tower of solutions can be obtained applying a lowering operator.

We also considered the role of quasinormal modes and frequencies in a Cauchy initial value problem. The results obtained from the initial value setup complement (and corroborate) the algebraic techniques used in the present work. In this context, we showed that there are no late-time tails and the time evolution is always stable. Therefore, at late enough times, an arbitrary solution can be written as a superposition of quasinormal modes. It follows that the quasinormal modes of the generalized P\"{o}schl-Teller potentials completely characterize the field dynamics. These results suggest that the rigorous analysis presented in \cite{Beyer:1998nu} for the properties of the usual P\"{o}schl-Teller quasinormal modes might be extended to our more general setup.

Although the focus of this paper is not on specific physical scenarios, applications of the results in perturbative General Relativity are straightforward. For instance, field evolution in pure de Sitter and anti-de Sitter spacetimes, and also in the near extreme de Sitter black holes and wormholes are described by generalized P\"{o}schl-Teller potentials. We comment some of those concrete scenarios in the appendices.
Nevertheless, the full role of the underling hidden algebra in the perturbative dynamics and its connection with the background geometry are not clear. 
Another interesting point is the completeness properties of the quasinormal mode spectra. The quasinormal modes of the usual P\"{o}schl-Teller potential are not stricto sensu complete \cite{Beyer:1998nu}. For the generalized forms of the P\"{o}schl-Teller potential considered in the present work, the analysis is more involving.
Those issues are being currently investigated.

\begin{acknowledgments}

A.~F.~C. acknowledges the support of FAPESP (grant No. 2013/07414-5), Brazil.
C.~M. acknowledges the support of FAPESP (grant No. 2015/24380-2) and CNPq (grant No. 307709/2015-9), Brazil.

\end{acknowledgments}

\section*{Journal notice}

This is an author-created, un-copyedited version of an article published in Classical and Quantum Gravity. IOP Publishing Ltd is not responsible for any errors or omissions in this version of the manuscript or any version derived from it. The Version of Record is available online at \url{https://doi.org/10.1088/1361-6382/aa9428}.

\newpage

\appendix

\section{Perturbative Dynamics on de Sitter spacetime}
 
Considering a physical system modeled by a generalized form of the P\"{o}schl-Teller potential, we will discuss quasinormal modes of pure de Sitter spacetime. 
In general, for a static and spherically symmetric spacetime 
there exists a coordinate system $(t,r,\theta,\phi)$ such that 
the metric can be expressed as
\begin{equation}\label{static_metric}
ds^{2} = -A(r)dt^2 + \frac{1}{B(r)}dr^2 + r^2d\Omega^{2} \,,
\end{equation} 
where $d\Omega^{2}$ denotes the line element on the $2$-sphere.
For simplicity, we consider the dynamics of a massless scalar field $\Phi$
in the Sitter spacetime satisfying the massless Klein-Gordon equation:
\begin{equation}\label{KG-equation}
\Box \Phi = 0  \, .
\end{equation}

After a decomposition in spherical harmonics of the form
\begin{equation}\label{expansion}
\Phi(t,r,\theta,\phi) = \sum_{\ell,m} \frac{1}{r}{\Psi_{\ell}(t,r)}Y_{\ell m}(\theta,\phi) \,,
\end{equation}
it is obtained the following differential equation,
\begin{equation}\label{Radialequation}
-\frac{\partial^{2}}{\partial{t}^{2}}\Psi_{\ell}(t,r) + \sqrt{A(r)B(r)}\frac{\partial}{\partial{r}}\left({\sqrt{A(r)B(r)}}\frac{\partial}{\partial{r}}\Psi_{\ell}(t,r)\right)  = V_{sc}(r)\Psi_{\ell}(t,r),
\end{equation}
where $V_{sc}(r)$ is the effective potential of the scalar perturbation,
depending solely on the metric coefficients in equation~(\ref{static_metric}): 
\begin{equation}\label{EffectivePotential}
 V_{sc}(r) 
 = \frac{\ell(\ell+1)}{r^{2}}{A(r)}
 + \frac{1}{2r}\left[ A'(r)B(r) + A(r)B'(r)\right] \,.
\end{equation}

For pure de Sitter spacetime, the metric coefficients in equation~(\ref{static_metric}) are 
\begin{equation}\label{de Sitter}
A(r) = B(r) = 1 - \frac{r^2}{l^2}\,,
\end{equation}
where $l$ is the de Sitter radius, related to a positive cosmological constant $\Lambda$ by $\Lambda = 3/l^2$.
With a change of variables $x = l\tanh^{-1}(r/l)$, where $x$ is the tortoise coordinate defined by $dx/dr = 1/\sqrt{A(r)B(r)}$, equation~(\ref{Radialequation}) reduces to a differential equation of the form~(\ref{maineq}), with the effective potential given by
\begin{equation}
V(x) = \frac{1}{l^2} \left[\frac{\ell(\ell+1)}{\sinh^{2}(r/l)} - \frac{2}{\cosh^{2}(r/l)} \right] \, .
\label{Poschl_teller-deSitter}
\end{equation}
Thus, $V_{+} = -2/l^2$ and $V_{-} = \ell(\ell+1)/l^2$. Following equation~(\ref{omega0_full_1}),
the fundamental quasinormal frequency is written as
\begin{equation}
\omega_{0} = - 
\frac{i}{l}\left( \frac{3}{2} + \ell \mp \frac{3}{2}\right) \, ,
\end{equation}
which is purely imaginary, indicating that for pure de Sitter spacetime the decay of a scalar field 
is non-oscillatory. This result can be verified with the fundamental frequency obtained in 
\cite{Du:2004jt} and \cite{lopez2006quasinormal} for the particular case of a massless scalar field
in $4$-dimensional de Sitter spacetime.

\section{Near extremal Schwarzschild-de Sitter spacetime}

Another relevant scenario in gravitational physics that can be represented by the dynamics of the P\"{o}schl-Teller potential is the Schwarzschild-de Sitter black hole, representing a black hole of mass $M$ and a cosmological constant $\Lambda = 3/l^2$. In this case, the metric has the form~(\ref{static_metric}) with
\begin{equation}\label{Schwarzschild-de-Sitter}
A(r) = B(r) = 1 - \frac{2M}{r} - \frac{r^2}{l^2}\,.  
\end{equation}
The Schwarzschild-de Sitter geometry possesses both an event horizon $r=r_{1}$ 
and a cosmological horizon $r=r_{2}$, given by the positive roots of 
\eqref{Schwarzschild-de-Sitter}.

We consider the near extremal limit of equation~(\ref{Schwarzschild-de-Sitter}), 
in which both horizons become arbitrarily close. In that case the 
metric element is approximated by \cite{Cardoso:2003sw,Molina:2003ff}
\begin{equation}\label{A(r)nearextremalSdS}
A(r) = \frac{2\kappa_{1}}{r_{2} - r_{1}}(r_{2} - r)(r - r_{1}) 
\,,
\end{equation}
where $\kappa_{1}$ is the surface gravity of the event horizon, 
$\kappa_{1} = \frac{1}{2}\vert dA(r)/dr \vert_{r_1}$. 
In this limit, the effective potential~(\ref{EffectivePotential}) of a scalar perturbation is approximated by:
\begin{equation}
 V(r) = \left[\frac{\ell(\ell+1)}{r_{1}^{2}}\right]
 \frac{2\kappa_{1}(r_{2} - r)(r - r_{1})}{r_{2} - r_{1}}\,.
 \label{potential_SdS}
\end{equation}
Similar expressions are obtained for electromagnetic and gravitational perturbations.

With a change of coordinates given by
\begin{equation}\label{tortoiseSdS}
x = \frac{1}{2\kappa_{1}}\ln \left(\frac{r - r_{1}}{r_{2} - r} \right)\,,
\end{equation}
it is obtained that the effective potential for the near extremal Schwarzschild de Sitter spacetime \eqref{potential_SdS} 
is approximated by the P\"{o}schl-Teller potential as introduced in equation~(\ref{poschl-teller_def})
with $\kappa = \kappa_{1}$ and the constant $V_{+}$ given by
\begin{equation}
 V_{+} = \frac{\ell(\ell+1)}{r_{1}^{2}} \,
 \frac{\kappa_{1}(r_{2} - r_{1})}{2} \,.
\end{equation}
Therefore, the associated quasinormal frequencies are those obtained from equation~(\ref{qnf}),
\begin{equation}
\omega_{n} =  \kappa_{1} \left[ - i\left(n + \frac{1}{2}\right) 
\pm 
\sqrt{
\frac{\ell(\ell+1) (r_{2} - r_{1})}{2 \kappa_{1} r_{1}^{2}} \,
- \frac{1}{4}} \right] \, , \,\,
n = 0,1,2\ldots \, .
\end{equation}

\end{document}